\documentclass[twocolumn,showpacs,preprintnumbers,
amsmath,amssymb,aps,nofootinbib,prl,superscriptaddress,
eqsecnum]{revtex4}
\usepackage{graphicx}

\makeatletter
\renewcommand{\p@subsection}{}
\makeatother

\begin{document}

\title{A novel spectral broadening from
vector--axial-vector mixing in dense matter}

\author{Masayasu Harada}
\affiliation{%
Department of Physics, Nagoya University,
Nagoya, 464-8602, Japan}
\author{Chihiro Sasaki}
\affiliation{%
Physik-Department,
Technische Universit\"{a}t M\"{u}nchen,
D-85747 Garching, Germany
}

\date{\today}

\begin{abstract}
The presence of baryonic matter leads to the mixing
between transverse $\rho$ and $a_1$ mesons 
through a set of $\omega\rho a_1$-type interactions,
which results in
the modification to the dispersion relation.
We show that a clear enhancement of the
vector spectral function appears below $\sqrt{s}=m_\rho$
for small three-momenta of the $\rho$ meson, 
and thus the vector spectrum exhibits broadening.
We also discuss its relevance to dilepton measurements.
\end{abstract}


\maketitle

%%%%%%%%%%%%%%%%%%%%%%%%%%%%%%%%%%%%%%%%%%%%%%%%%%%%%
%{\it Introduction}\,---
%%%%%%%%%%%%%%%%%%%%%%%%%%%%%%%%%%%%%%%%%%%%%%%%%%%%
In-medium modifications of hadrons have been extensively
explored in the context of chiral dynamics of QCD~\cite{review,rapp}.
Due to an interaction with pions in the heat bath, the vector 
and axial-vector current correlators are mixed.
At low temperatures or densities a low-energy theorem based on 
chiral symmetry describes this mixing (V-A mixing)~\cite{theorem}.
The effects to the thermal vector spectral function have been
studied through the theorem~\cite{vamix}, or using chiral reduction
formulas based on a virial expansion~\cite{chreduction},
and near critical temperature in a chiral effective field theory 
involving the vector and axial-vector mesons as well as the 
pion~\cite{our}.

It has been derived, as a novel effect at finite baryon density,
that a Chern-Simons term leads to mixing between the vector and 
axial-vector fields in a holographic QCD model~\cite{hqcd}.
This mixing modifies the dispersion relation of the transverse
polarizations and will affect the in-medium current correlation
functions independently of specific model dynamics.
In this letter we study the effect of the vector--axial-vector
mixing to the in-medium spectral functions which is the main input 
to the experimental observables.
We show that 
the mixing produces
a clear enhancement of the vector spectral function 
which appears below $\sqrt{s}=m_\rho$,
and that
the vector spectral function 
is broadened due to the mixing.
We will discuss its relevance to dilepton measurements.

\begin{figure}
\begin{center}
\includegraphics[width=8cm]{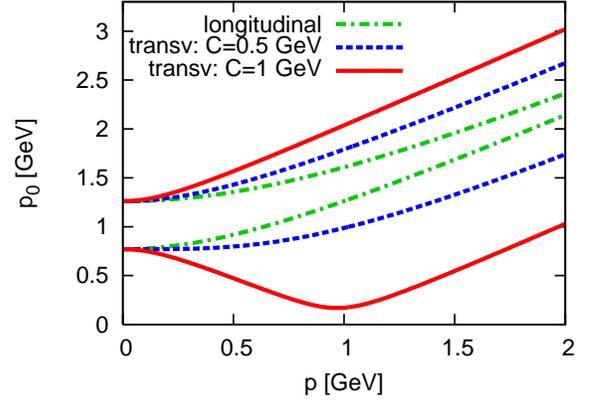}
\caption{
The dispersion relation of the $\rho$ (lower 3 curves)
and $a_1$ (upper 3 curves) mesons for $C=0.5$, and $1$ GeV.
}
\label{dispersion}
\end{center}
\end{figure}

%%%%%%%%%%%%%%%%%%%%%%%%%%%%%%%%%%%%%%%%%%%%%%%%%%%%
%{\it V-A mixing term}\,---
%%%%%%%%%%%%%%%%%%%%%%%%%%%%%%%%%%%%%%%%%%%%%%%%%%%
At finite baryon density a system preserves parity 
but violates charge conjugation invariance.
Chiral Lagrangians thus in general build 
in the term
\begin{equation}
{\cal L}_{\rho a_1} = 2C\,\epsilon^{0\nu\lambda\sigma}
\mbox{tr}\left[ \partial_\nu V_\lambda \cdot A_\sigma
{}+ \partial_\nu A_\lambda \cdot V_\sigma \right]\,.
\label{vaterm}
\end{equation}
This mixing results in the dispersion relation~\cite{hqcd}
\begin{equation}
p_0^2 - \bar{p}^2 = \frac{1}{2}\left[ 
m_\rho^2 + m_{a_1}^2 \pm \sqrt{(m_{a_1}^2 - m_\rho^2)^2
{}+ 16 C^2 \bar{p}^2}
\right]\,,
\label{disp}
\end{equation}
which describes the propagation of a mixture of the transverse $\rho$ 
and $a_1$ mesons with non-vanishing three-momentum $|\vec{p}|=\bar{p}$.
The longitudinal polarizations, on the other hand,
follow the standard dispersion
relation, $p_0^2 - \bar{p}^2 = m_{\rho,a_1}^2$.
When the mixing vanishes as $\bar{p} \to 0$, Eq.~(\ref{disp}) with
lower sign provides $p_0 = m_\rho$ and it with upper sign
does $p_0 = m_{a_1}$.  In the following, we call the mode following
the dispersion relation with the lower sign in Eq.~(\ref{disp})
``the $\rho$ meson'', and it with the upper sign ``the $a_1$ meson''.
In a holographic QCD approach the coupling $C$ depends on the 
baryon density $n_B$ and is found $C \simeq 1\,\mbox{GeV}\cdot(n_B/n_0)$ 
with normal nuclear matter density $n_0 = 0.16$ fm$^{-3}$~\cite{hqcd}.
Figure~\ref{dispersion} shows the dispersion relation~(\ref{disp}).
%%%%%%%%%%%%%%%%%%%%%%%%%%%%%%%%%%%%%%%%%%%%%
%%%%%%%%%%%%%%%%%%%%%%%%%%%%%%%%%%%%%%%%%%%
For very large $\bar{p}$ the longitudinal and transverse 
dispersions are in parallel with a finite gap, $\pm C$. 
The dispersion relation (\ref{disp}) also indicates a possibility
of vector condensation for a large $C$~\cite{hqcd}.

%%%%%%%%%%%%%%%%%%%%%%%%%%%%%%%%%%%%%%%%%%%%%%%%%%%%%%%%%%%
%{\it Vector spectral function and dilepton rates}\,---
%%%%%%%%%%%%%%%%%%%%%%%%%%%%%%%%%%%%%%%%%%%%%%%%%%%%%%%%%%%%%%

%%%%%%%%%%%%%%%%%%%%%%%%%%%%%%%%%%%%%%%%%%%%%
\begin{figure*}
\begin{center}
\includegraphics[width=8.5cm]{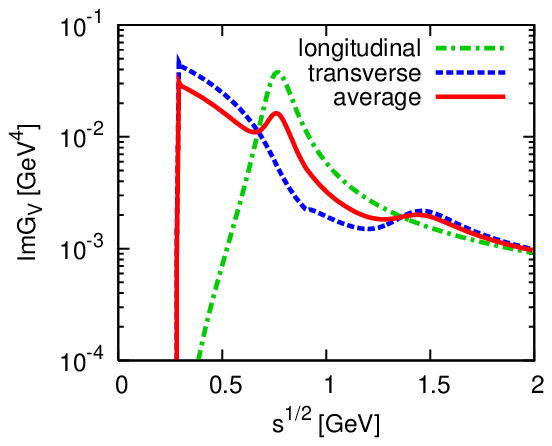}
\includegraphics[width=8.5cm]{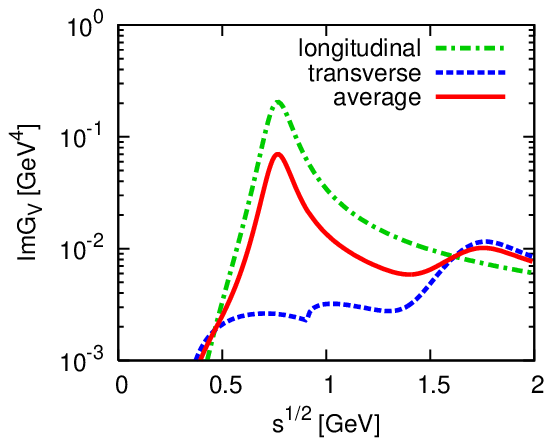}
\caption{
The vector spectral function for $C=1$ GeV.
The curves of the left figure are calculated integrating over 
$0 < \bar{p} < 0.5$ GeV, and those of the right figure
over $0.5 < \bar{p} < 1$ GeV. 
}
\label{CC1}
\end{center}
\end{figure*}
%%%%%%%%%%%%%%%%%%%%%%%%%%%%%%%%%%%%%%%%%%%

The vector-current correlation function in matter is decomposed 
into the longitudinal and transverse parts as
\begin{equation}
G_V^{\mu\nu}(p_0,\vec{p})
= P_L^{\mu\nu} G_V^L(p_0,\vec{p}) 
{}+ P_T^{\mu\nu} G_V^T(p_0,\vec{p})\,,
\label{ccv}
\end{equation}
with the polarization tensors $P_{L,T}^{\mu\nu}$
and momentum $p^\mu = (p_0,\vec{p})$.
Using the bare propagator inverse,
$D_{V,A} = s - m_{\rho,a_1}^2 + im_{\rho,a_1}\Gamma_{\rho,a_1}(s)$,
$G_V^L$ and $G_V^T$ are expressed as
\begin{equation}
G_V^L = \left(\frac{g_\rho}{m_\rho}\right)^2
\frac{-s}{D_V}\,,
\quad
G_V^T = \left(\frac{g_\rho}{m_\rho}\right)^2
\frac{-s D_A + 4C^2\bar{p}^2}
{D_V D_A - 4 C^2 \bar{p}^2}\,,
\label{LT}
\end{equation}
with $s=p^2$ being the squared four-momentum and $g_\rho$
the coupling strength of the $\rho$ meson to the vector current.
We have imposed gauge invariance on the vector current to get the 
form (\ref{LT}).
The spin-averaged correlator is given by
$G_V = \frac{1}{3}\left[ G_V^L + 2 G_V^T \right]$.
The vector spectral function is defined as the imaginary part
of the vector correlator~(\ref{ccv}).
We define the integrated spectrum over three momentum by
\begin{equation}
\mbox{Im}G_V(s) = \int\frac{d^3\vec{p}}{2p_0}
\mbox{Im}G_V(p_0,\vec{p})\,.
\end{equation}
Equation~(\ref{LT}) indicates that the mixing at finite three 
momentum $\bar{p}$ affects the real part of the transverse $\rho$ 
self-energy.
We use the vacuum decay widths $\Gamma$ to illustrate its influence 
over the spectrum although $\Gamma$ in dense matter are considered 
to be broadened~\cite{rapp}. 
We take the following experimental values for further calculations:
$m_\pi = 0.14$ GeV, $m_\rho = 0.77$ GeV, $m_{a_1} = 1.26$
$g_\rho = 0.119$ GeV$^2$, 
$\Gamma_{\rho}(s=m_\rho^2) = 0.15$ GeV~\cite{pdg}.
For the $a_1$ decay width we use 
$\Gamma_{a_1}(s=m_{a_1}^2) = 0.3$ GeV as a typical example.

We show the vector spectral function in Fig.~\ref{CC1}.
The transverse spectrum presents two bumps due to the mixing: 
the lower one corresponds to the $\rho$
whose mass is shifted downward, and the upper one to the $a_1$ 
whose mass is shifted upward in compared with the longitudinal 
polarizations (see Fig.~\ref{dispersion}). Since two pion
annihilation is assumed to be dominant in the $\rho$ meson 
decay, the contribution at low $\sqrt{s}$ is cut off at
threshold $\sqrt{s}=2m_\pi$.
Figure~~\ref{CC1} (left) shows a clear enhancement of the spectrum 
below $\sqrt{s}=m_\rho$ due to the mixing.
This enhancement becomes much 
suppressed when the $\rho$ meson is moving with a large three-momentum
as shown in Fig.~\ref{CC1} (right). The upper
bump now emerges more remarkably and becomes a clear indication
of the in-medium effect from the $a_1$ via the mixing.
The presence of the two bumps in the transverse part
leads to some broadening of the spin-averaged spectrum.

For more realistic evaluations one needs to include nuclear 
many-body dynamics into meson decay widths.
This will be another source of in-medium broadening and eventually 
the vector correlator may not exhibit a clear maximum. 
Besides the iso-vector $\rho$-$a_1$ mesons, 
the mixing between 
the $\omega$ and $f_1(1285)$ mesons as well as that between
the $\phi$ and $f_1(1420)$ mesons
in iso-scalar channel also exists 
and changes the dispersion relations. 
This is controlled by the same mixing strength $C$ which can
be smaller in three-color QCD than the value predicted 
in holographic QCD models. 
In such a case the spectrum
enhancement in low $\sqrt{s}$ region becomes more moderate 
but the effect is still relevant 
to the vector spectrum of the $\rho$ mesons carrying large $\bar{p}$.
As a result, 
the averaged spectrum might have a broad bump with its maximum
slightly shifted {\it downward} due to the mixing.
Thus, it is expected that those mixing have some relevance to explain
in-medium ``mass shift'' of the $\rho$, $\omega$ and $\phi$ mesons 
observed by CBELSA/TAPS and KEK-PS-E325~\cite{taps,kek}.

As an application of the above in-medium spectrum, we calculate
the production rate of a lepton pair emitted from dense 
matter through a decaying virtual photon.
The differential production rate in a medium for fixed 
temperature $T$ and baryon density $n_B$ is expressed in terms of 
the imaginary part of the vector current correlator as~\cite{rapp}
\begin{equation}
\frac{dN}{d^4p}(p_0,\vec{p};T,n_B)
=\frac{\alpha^2}{\pi^3 s}\frac{1}{e^{p_0/T}-1}
\mbox{Im}G_V (p_0,\vec{p};T,n_B)\,,
\end{equation}
where $\alpha = e^2/4\pi$ is the electromagnetic coupling constant.
The three-momentum integrated rate is given by
\begin{equation}
\frac{dN}{ds}(s;T,n_B) 
=\int\frac{d^3\vec{p}}{2p_0}\frac{dN}{d^4p}(p_0,\vec{p};T,n_B)\,.
\end{equation}

\begin{figure}
\begin{center}
\includegraphics[width=8.5cm]{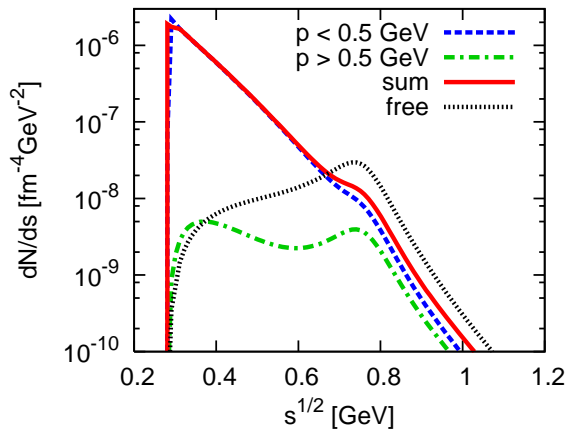}
\caption{
The dilepton production rate at $T=0.1$ GeV for $C=1$ GeV. 
Integration over $0 < \bar{p} < 0.5$ GeV (dashed) and 
$0.5 < \bar{p} < 1$ GeV (dashed-dotted) was carried out.
}
\label{rate}
\end{center}
\end{figure}

Figure~\ref{rate} presents the integrated rate at $T=0.1$ GeV
for $C=1$ GeV.
%%%%%%%%%%%%%%%%%%%%%%%%%%%%%%%%%%%%%%%%%%%%%
%%%%%%%%%%%%%%%%%%%%%%%%%%%%%%%%%%%%%%%%%%%
One clearly observes a strong three-momentum dependence and an 
enhancement below $\sqrt{s}=m_\rho$ due to the Boltzmann distribution 
function which result in a strong spectral broadening.
The total rate is mostly governed by the spectrum with 
low momenta $\bar{p}<0.5$ GeV due to the large mixing parameter $C$.
When density is decreased, 
the mixing effect gets irrelevant and
consequently in-medium effect in low $\sqrt{s}$ region is reduced 
in compared with that at higher density.  
The calculation performed in hadronic many-body theory in fact
shows that the $\rho$ spectral function with a low momentum
carries details of medium modifications~\cite{riek}.
One may have a chance to observe it in heavy-ion collisions 
with certain low-momentum binning at J-PARC, GSI/FAIR and RHIC 
low-energy running.

\begin{figure}
\begin{center}
\includegraphics[width=8.5cm]{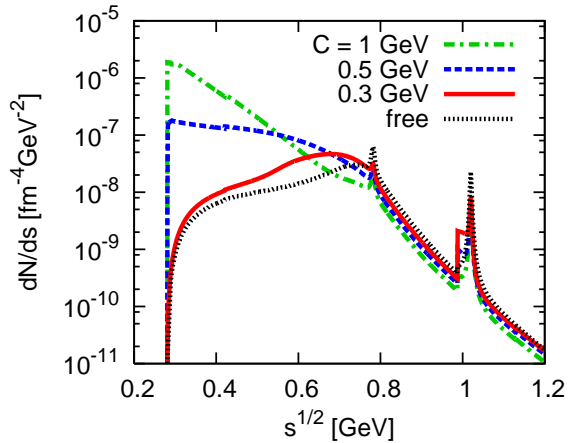}
\caption{
The dilepton production rate at $T=0.1$ GeV
with various mixing strength $C$.
Integration over $0 < \bar{p} < 1$ GeV was done.
Free decay widths are used.
}
\label{rateV}
\end{center}
\end{figure}

It is straightforward to introduce other V-A mixing between 
$\omega$-$f_1(1285)$ and $\phi$-$f_1(1420)$.
We use the constant widths of narrow peaked mesons above threshold:
$\Gamma_\omega = 8.49$ MeV, $\Gamma_\phi = 4.26$ MeV,
$\Gamma_{f_1(1285)}=24.3$ MeV and $\Gamma_{f_1(1420)}=54.9$ 
MeV~\cite{pdg}.
The coupling constants of $\omega$ and $\phi$ mesons to the vector 
current are given by
\begin{equation}
g_\omega = \frac{1}{3}\frac{m_\omega^2}{m_\rho^2}\,g_\rho\,,
\quad
g_\phi = \frac{\sqrt{2}}{3}\frac{m_\phi^2}{m_\rho^2}\,g_\rho\,.
\end{equation}
Figure~\ref{rateV} shows the integrated rate at $T=0.1$ GeV
with several mixing strength $C$ which are phenomenological option.
%%%%%%%%%%%%%%%%%%%%%%%%%%%%%%%%%%%%%%%%%%%%%
%%%%%%%%%%%%%%%%%%%%%%%%%%%%%%%%%%%%%%%%%%%
One observes that the enhancement below $m_\rho$ is suppressed
with decreasing mixing strength. This forms into a broad bump
in low $\sqrt{s}$ region and its maximum moves toward $m_\rho$.
Similarly, some contributions are seen just below $m_\phi$.
This effect starts at threshold $\sqrt{s}= 2 m_K$ in the present
analysis because of 
$\Gamma_\phi(s) = \Theta(s-4m_K^2)\Gamma_\phi(m_\phi)$.
Self-consistent calculations of the spectrum in dense medium
will provide a smooth change and this eventually makes the $\phi$
meson peak somewhat broadened.

%%%%%%%%%%%%%%%%%%%%%%%%%%%%%%%%%%%%%%%%%%%%%%%%%%%%%%%
%{\it Conclusions}\,---
%%%%%%%%%%%%%%%%%%%%%%%%%%%%%%%%%%%%%%%%%%%%%%%%%%%%%
The relevance of this mixing in dense matter essentially relies on
how the strength $C$ is precisely determined. Holographic QCD 
approach predicts a strong mixing. However, the models based on 
the gravity-gauge correspondence are formulated in large $N_c$ limit. 
Their prediction of observables may have a non-negligible $1/N_c$ 
correction~\cite{matsuzaki}. This suggests a possibility that $C$ is 
smaller in realistic QCD.
One might consider to replace the mixing term (\ref{vaterm}) 
with the $\omega$-$\rho$-$a_1$ term which has been shown to arise from 
the gauged Wess-Zumino-Wittem term in chiral Lagrangians~\cite{kaiser} 
or alternatively from the reduction of five-dimensional Chern-Simons 
term to four dimensions~\cite{hill},
\begin{equation}
{\cal L}_{\omega\rho a_1} = g_{\omega\rho a_1}
\langle \omega_0\rangle \epsilon^{0\nu\lambda\sigma}
\mbox{tr}\left[ \partial_\nu V_\lambda \cdot A_\sigma
{}+ \partial_\nu A_\lambda \cdot V_\sigma \right]\,,
\end{equation}
where the $\omega$ field is replaced with its expectation value
given by $\langle \omega_0\rangle = g_{\omega NN}\cdot n_B/m_\omega^2$.
One finds with empirical numbers
$C = g_{\omega\rho a_1}\langle \omega_0\rangle \simeq 0.1$ GeV
at normal nuclear matter density. This relatively much weaker mixing
has little importance in the correlation functions.
It is plausible to assume an actual value of $C$ in QCD
in the range $0.1 < C < 1$ GeV since the strong mixing in holographic
QCD models contains higher members of Kaluza-Klein (KK) modes other
than the lowest $\omega$ meson
and those higher excitations are embedded in $C$.
Some importance of the higher KK modes {\it even in vacuum} in the 
context of holographic QCD can be seen in the pion electromagnetic 
form factor at the photon on-shell:
This is saturated by the lowest four vector mesons in a top-down 
holographic QCD model~\cite{ss,matsuzaki2}. 
In hot and dense environment those higher members get modified
and the masses might be somewhat decreasing evidenced in an
in-medium holographic model~\cite{sstem}. This might provide
a strong V-A mixing $C > 0.1$ GeV in three-color QCD
and the dilepton measurements may be a good testing ground.

It is also an interesting issue to address a change of the vector 
correlator with the V-A mixing toward chiral symmetry restoration. 
The mixing (\ref{vaterm}) is chirally symmetric and thus 
does not approach zero toward the chiral restoration
in contrast to the vanishing V-A mixing near the critical 
temperature $T_c$ without baryon density~\cite{our}.
A spontaneous breaking of Lorentz invariance via the omega 
condensation could increase the mixing strength $C$ near chiral 
restoration~\cite{isb}.
Furthermore,
if meson masses drop due to partial restoration of chiral
symmetry assuming a second- or weak first-order transition
in high baryon density but low temperature region,
the ground state near the critical point may favor
vector condensation even for a moderate mixing strength.
This will be reported elsewhere.

%%%%%%%%%%%%%%%%%%%%%%%%%%%%%%%%%%%%%%%%%%%%%%%%%%%%%%%%
\subsection*{Acknowledgments}
%%%%%%%%%%%%%%%%%%%%%%%%%%%%%%%%%%%%%%%%%%%%%%%%%%%%%%

We acknowledge stimulating discussions with B.~Friman,
N.~Kaiser, S.~Matsuzaki, M.~Rho and W.~Weise.
The work of C.S. has been supported in part 
by the DFG cluster of excellence ``Origin and Structure of the 
Universe''.
The work of M.H. has been supported in part by
the JSPS Grant-in-Aid for Scientific Research (c) 20540262
and Global COE Program 
``Quest for Fundamental Principles in the Universe''
of Nagoya University provided by Japan Society for the
Promotion of Science (G07).

%%%%%%%%%%%%%%%%%%%%%%%%%%%%%%%%%%%%%%%%%%%%%%%%%%%%%%%%
%%%%%%%%%%%%%%%%%%%%%%%%%%%%%%%%%%%%%%%%%%%%%%%%%%%%%%%%%

\end{document}